# Hf/Zr Superlattice-Based High-κ Gate Dielectrics with Dipole Layer Engineering for Advanced CMOS

**AUTHOR NAMES**

Taeyoung Song,[1,4,*] Sanghyun Kang,[1,2,4,*], Yu Hsin Kuo[1], Jiayi Chen[1], Lance Fernandes[1], Nashrah Afroze[1], Mengkun Tian[1], Hyoung Won Baac[2], Changhwan Shin[3] and Asif Islam Khan[1]

**AUTHOR ADDRESS**

[1]School of Electrical and Computer Engineering, Georgia Institute of Technology, Atlanta, GA, 30332, USA

[2]Department of Electrical and Computer Engineering, Sungkyunkwan University, Suwon, 16419, Republic of Korea

[3]School of Electrical Engineering, Korea University, Seoul, 02841, Republic of Korea

[4]These authors contributed equally

**KEYWORDS**

Ultrathin high-κ, Superlattice, dipole engineering, reliability, RMG process, GAAFET

**ABSTRACT**

Advanced logic transistors require gate dielectrics that achieve sub-nanometer equivalent oxide thickness (EOT), suppress leakage, and satisfy three key requirements: (i) compatibility with RMG-like high-temperature processing, (ii) sufficient $V_{th}$ tunability for multi-$V_{th}$ design, and (iii) high device reliability. However, meeting all of these requirements at once has been difficult with conventional high-κ systems. In this work, we demonstrate that our Hf/Zr-based gate stacks quantitatively satisfy these conditions. (i) After a 700 °C $N_2$ anneal, the HZH superlattice achieves EOT = 7.3 Å, lower than conventional $HfO_2$-only stacks (8.5 Å) while maintaining comparable leakage. (ii) Embedding a 3 Å $Al_2O_3$ dipole within the $HfO_2/ZrO_2/HfO_2$ superlattice (HZHA) breaks the conventional dipole trade-off, achieving an 8.4 Å EOT—lower than the 9.0 Å of a standard $HfO_2/Al_2O_3$ stack—while providing a >200 mV $V_{FB}$ shift, thereby enabling multi-Vth tuning without compromising scaling. (iii) Furthermore, under −2 V negative-bias temperature stress at 125 °C for 100 s, HZHA and HA exhibit comparable $V_{FB}$ drifts of 87 mV and 97 mV, respectively confirming that strong $V_{th}$ tunability and sub-nanometer EOT can be achieved without compromising stability. In addition to these quantitative advances, this study reveals previously unreported physical insights into dipole behavior and interfacial diffusion in ultrathin Hf/Zr multilayers. These results establish HZHA as an RMG-compatible, $V_{th}$-tunable, low-EOT dielectric platform capable of supporting logic scaling beyond the 1 nm frontier.

**INTRODUCTION**

The demand for more powerful and energy-efficient computing, driven by applications in artificial intelligence (AI), high-performance computing, and edge devices, has accelerated the miniaturization of logic transistors. In accordance with Moore's Law, continued scaling to smaller device dimensions improves computational efficiency and integration density, but also presents new challenges in device design and materials integration. Gate-all-around (GAA) field-effect transistors (FETs) have emerged as a key solution to maintain electrostatic control as transistor width is reduced.[1-6] Compared to traditional FinFETs, GAA transistors offer superior electrostatic control and reduced short-channel effects, making them well-suited for continued scaling (Figure 1A). By completely wrapping the gate around the channel, GAA FETs suppress short-channel effects and improve performance over traditional FinFETs, enabling further scaling of CMOS technology. However, the relentless drive toward nanometer-scale dimensions imposes severe constraints on integrating critical materials—such as work function metals and gate dielectrics—into an ever-shrinking volume (Figure 1B). In particular, gate dielectrics must provide extremely high capacitance in an ultrathin layer while also allowing tunable threshold voltages ($V_{th}$) to meet performance requirements. Stronger gate electrostatic control is needed to mitigate short-channel effects, and higher gate capacitance is

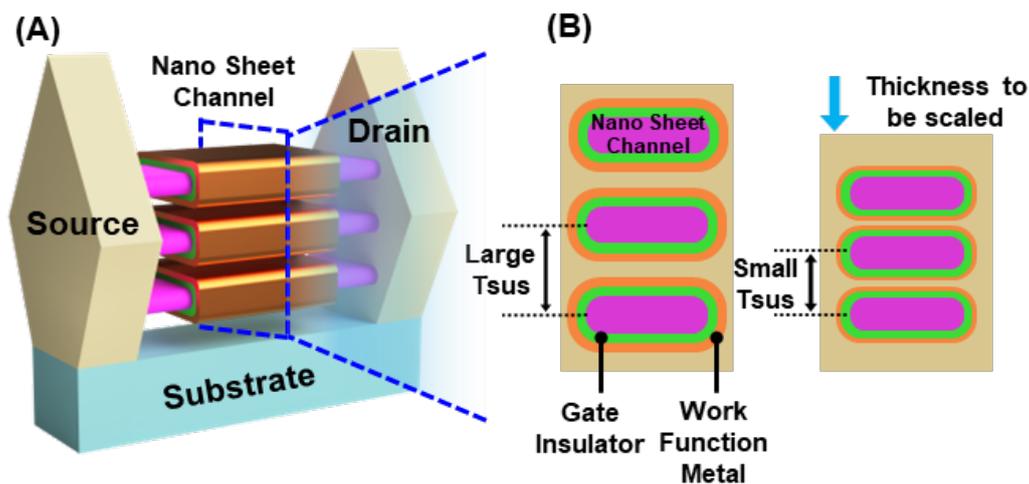

required to sustain drive current for high-speed operation.[7-12]

**Figure 1. The Need for Ultra-Thin Gate Oxides with High-k Dielectrics and Low Gate Leakage** (A) 3D schematic of a GAAFET structure. (B) Cross-sectional view of stacked nanosheets in a GAAFET, highlighting the increasingly constrained space for gate insulators and work function metals in advanced logic transistors.

One approach to achieve high capacitance at reduced thickness is the integration of advanced high-κ dielectric materials. Recent studies have explored laminated and superlattice oxides based on $HfO_2$, $ZrO_2$, and Hf–Zr mixed oxides (HZO), particularly near the morphotropic phase boundary (MPB), to enable further EOT scaling.[13-19] While many of these efforts have focused on metal–insulator–metal capacitors for ferroelectric memory or DRAM applications and sometimes metal–oxide–semiconductor (MOS) structures, high-temperature annealing steps are often omitted, raising concerns about thermal stability. In practical CMOS integration, gate dielectrics must withstand subsequent high-temperature processes such as dopant activation or defect annealing without degradation to ensure long-term device reliability. Post-deposition annealing (PDA) is critical for passivating defects and improving material properties; inadequate thermal treatment can leave the dielectric vulnerable to instability under operating conditions. As such, understanding the impact of high-temperature annealing on advanced high-κ gate stacks is essential for their integration into future technology nodes.[20, 21]

Another key challenge at aggressively scaled EOT is the increase in gate leakage current due to direct tunneling, which can lead to excessive power consumption. An intrinsic trade-off exists between a dielectric's permittivity (κ) and its bandgap. High-κ materials enable a reduction in gate oxide thickness, enhancing electrostatic gate control. However, their relatively low bandgap energies can lead to increased leakage currents under operating conditions.[22, 23] Therefore, optimizing the κ–bandgap balance is critical to minimizing leakage while maintaining gate capacitance, particularly in the sub-nanometer EOT regime.

Moreover, as physical dimensions shrink, incorporating work function metals to modulate the threshold voltage ($V_{th}$) becomes increasingly difficult due to limited gate volume. In this context, dipole engineering within the gate stack provides a "volume-less" method for $V_{th}$ tuning. By introducing ultrathin dielectric interlayers with differing oxygen areal densities

(typically at the high-κ/SiO$_2$ interface), dipoles which shift the flatband voltage ($V_{FB}$) without adding physical thickness can be formed. Various dipole materials, such as La$_2$O$_3$ and Al$_2$O$_3$, and their thickness dependence have been extensively studied.[24-26] However, systematic investigations into the optimal position of the dipole layer within the gate stack remain limited. Strategic placement of dipole layers could provide additional degrees of freedom for precise $V_{th}$ control in highly scaled devices.

In this study, we systematically investigate four gate dielectric stacks—HfO$_2$-only, HfO$_2$/ZrO$_2$/HfO$_2$ (HZH), ZrO$_2$/HfO$_2$/ZrO$_2$ (ZHZ), and HZO/ZrO$_2$/HZO (HZZ)—to evaluate their impact on EOT and gate leakage current. A high-temperature 700 °C PDA is found to markedly improve flatband voltage stability under negative bias stress (NBS), with the HZH and ZHZ configurations demonstrating significantly enhanced reliability compared to the conventional HfO$_2$-only baseline. Furthermore, we systematically explore dipole engineering by incorporating an ultrathin Al$_2$O$_3$ layer at various positions within the stack to modulate the threshold voltage. While such Al$_2$O$_3$ integration inevitably increases the EOT due to low dielectric constant of Al$_2$O$_3$, this drawback is mitigated by employing the HZH structure. Consequently, the optimized HZHA gate stack (HZH with Al$_2$O$_3$ dipole) achieves a favorable combination of reduced EOT, maintaining $V_{th}$ tunability, high-temperature process compatibility, and enhanced reliability, positioning it as a strong candidate for future CMOS logic devices. These findings highlight the HZHA gate stack as a promising platform for enabling EOT scaling, threshold voltage control, and long-term reliability in advanced CMOS logic technologies.

**RESULT AND DISCUSSION**

**Investigation of High-κ Gate Stack Structures for Enhanced Capacitance, Leakage Control and Reliability**

We first compared the electrical characteristics of three gate stack structures – $HfO_2$-only, HZH, and ZHZ – each with the same total thickness but different layer compositions. The MOS capacitors were fabricated following the process flow shown in Figure S1, with the total gate insulator thickness of 21 Å and additional details explained in the TEM and EDS images (Figure S2, S3). Figure 2A illustrates the layer configurations for the $HfO_2$ reference and the laminated HZH and ZHZ stacks with layer thicknesses in angstroms. Capacitance–voltage (C–V) and gate leakage current density–voltage ($J_g$–V) measurements were performed to evaluate the performance of these configurations (Figures 2B–G).

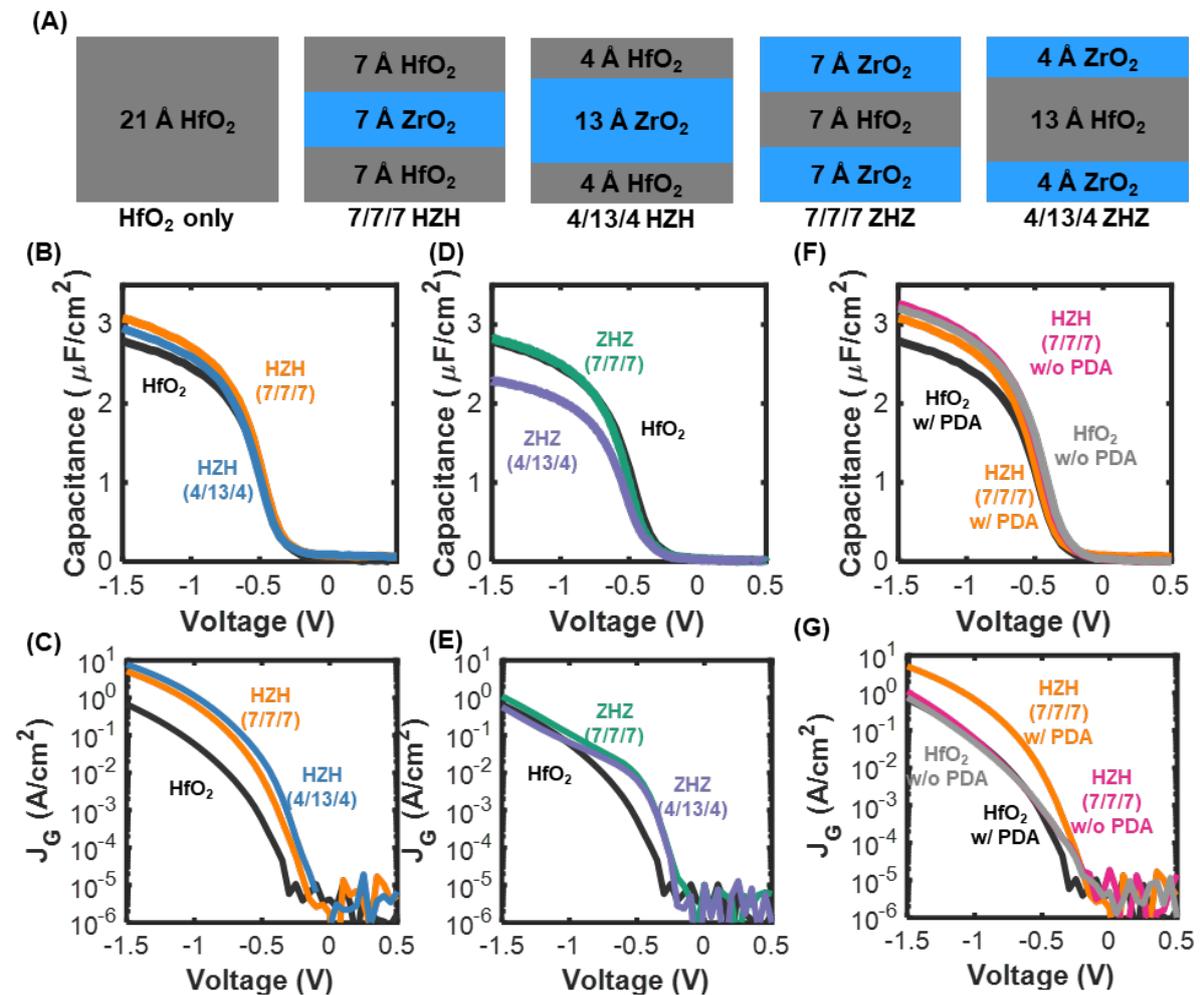

**Figure 2. Gate stack engineering for enhanced capacitance and minimized gate leakage current** (A) Schematic of the HfO$_2$-only, HZH, and ZHZ gate stack configurations with varying thicknesses (B) and (C) Capacitance–voltage (C–V) and gate leakage current–voltage (J$_g$–V) characteristics of HfO$_2$ and HZH gate stacks (D) and (E) Capacitance–voltage (C–V) and gate leakage current–voltage (J$_g$–V) characteristics of HfO$_2$ and ZHZ gate stacks (F) and (G) Capacitance–voltage (C–V) and gate leakage current–voltage (J$_g$–V) characteristics of HfO$_2$ and HZH stacks, both with and without post-deposition annealing (PDA).

This demonstrates that the HZH stack consistently exhibits a higher capacitance than the HfO$_2$-only sample, indicating an increased effective dielectric constant due to the incorporation of the intermediate ZrO$_2$ layer. In contrast, the ZHZ stack shows a capacitance that is comparable to or slightly lower than that of HfO$_2$-only, underscoring the strong influence of layer ordering on overall permittivity. Notably, the symmetric 7/7/7 Å HZH laminate achieved an EOT of 7.3 Å, a significant improvement over the 8.5 Å EOT of the HfO$_2$-only control. A thicker asymmetric HZH stack (4/13/4 Å) yielded a slightly higher EOT of 7.8 Å, confirming that the HfO$_2$:ZrO$_2$ thickness ratio impacts the dielectric performance. A similar trend is observed in the ZHZ configuration: the symmetric 7/7/7 Å ZHZ stack attained an EOT of 8.4 Å, whereas the less optimized 4/13/4 Å ZHZ stack showed a much larger EOT of 10.6 Å. These results highlight the importance of both the stacking sequence and individual layer thickness optimization in achieving the desired high-κ performance.

Significant research has been devoted to achieving low EOT in HfO$_2$- and ZrO$_2$-based dielectrics through compositional and structural engineering. Recent studies on ultra-thin ferroelectric HfO$_2$–ZrO$_2$ superlattice capacitors have shown that a coexistence of the ferroelectric orthorhombic (o) phase and the antiferroelectric tetragonal (t) phase can effectively reduce the EOT, leveraging negative capacitance or high permittivity effects at the phase boundary.[13, 16, 18] In our HZH laminate, X-ray diffraction (Figure S4) confirms the presence of mixed crystal phases: a dominant o-(111) peak at 2θ ≈ 30.4 ° (accounting for 78.13 % of the film) alongside a weaker t-(011) peak (21.25 %) and a minor monoclinic phase residual. This phase coexistence is a key factor contributing to the reduced EOT of the HZH stack compared to a single-layer $HfO_2$ film. The $ZrO_2$ incorporation helps stabilize the desirable orthorhombic phase of $HfO_2$, which has higher permittivity, thereby boosting capacitance for a given thickness.

Besides phase stabilization, the $HfO_2$:$ZrO_2$ thickness ratio is pivotal in optimizing dielectric performance. Our comparison of 7/7/7 Å versus 4/13/4 Å HZH stacks shows that the distribution of Hf and Zr dictates the o versus t phase balance and, in turn, the attainable EOT. Sufficient $HfO_2$ content is required to stabilize the orthorhombic phase; when the $ZrO_2$ fraction becomes excessive—as in the 4/13/4 Å stack—the structure favors the tetragonal phase, leading to a thicker effective oxide. Precise control of individual layer thicknesses is therefore essential to reproducibly target the desired phase composition and permittivity.

The gate leakage characteristics of the stacks are shown in Figure 2C and 2E. The HZH stack exhibits only a slight increase in leakage current compared to the $HfO_2$-only device, while the ZHZ stack maintains a similar leakage level to the $HfO_2$ reference. This suggests that inserting $ZrO_2$ layers can enhance capacitance without dramatically compromising leakage current. Leakage current densities in both HZH and ZHZ laminates remain comparable to, or only slightly higher than, those of pure $HfO_2$ across the operating range, consistent with their band edge alignment: $HfO_2$'s larger bandgap (~5.8 eV) provides the dominant barrier, while the modest introduction of $ZrO_2$ (~5.4 eV) only marginally increases tunneling probability, leaving overall leakage acceptable for logic applications.[22, 23] However, careful tuning of the $ZrO_2$ layer's thickness and position is required to balance the trade-off between higher κ and potential increases in leakage. The HZH stack shows a favorable trade-off: it achieves a substantially lower EOT (higher capacitance) than the HfO$_2$-only baseline, with only a minor leakage penalty, making it attractive for future high-performance devices. The ZHZ (7/7/7) stack, on the other hand, offers moderate capacitance (similar to HfO$_2$) but with stable leakage behavior, illustrating how different laminate designs can target either maximum capacitance or minimized leakage according to application needs.

To assess the influence of high temperature processing, we subjected each gate stack to a 700 °C PDA for 30 s in N$_2$ and compared the results with unannealed controls. Without PDA, the HfO$_2$ only stack exhibited an exceptionally low EOT of 7.1 Å and low gate leakage. Following the 700 °C anneal, its EOT increased to 8.5 Å and the leakage current rose by roughly 50 % (Figure 2F, G). The HZH stack displayed a more modest EOT shift—from 7.0 Å to 7.3 Å—but experienced a proportionally larger rise in leakage current.

A notable observation in our study is that both EOT and gate leakage current density ($J_g$) increased after high-temperature annealing even for conventional HfO$_2$ sample. As 700°C 30s annealing process isn't that high heat budget process enough to transit the phase of crystalline of HfO$_2$, the primary cause of this degradation is the formation of interfacial silicates at the SiO$_2$/high-κ interface during annealing. At elevated temperatures, HfO$_2$ or ZrO$_2$ in contact with the interfacial SiO$_2$ layer can react to form hafnium silicate (HfSiO$_4$) or zirconium silicate (ZrSiO$_4$). While these silicate phases are thermally stable, they possess a significantly lower dielectric constant than pure HfO$_2$ or ZrO$_2$, resulting in an effective thickness increase. Indeed, silicate formation is a well-known limiting factor for EOT scaling in high-κ dielectrics.[27] Meanwhile, prior to PDA, the as-deposited EOT values exhibited similar value with best value of HfO$_2$–ZrO$_2$ systems from the previous reported paper.[13] However, after PDA HZH sample would have another effect which is intermixing of HZH and interaction within interface between HfO$_2$ and ZrO$_2$ which leads to more increase in the gate leakage.

In our stacks, the ZrO$_2$-containing samples were slightly more susceptible to silicate-induced EOT increases than the HfO$_2$-only sample. This is expected because ZrSiO$_4$ formation occurs at a lower temperature than HfSiO$_4$.[27] Thus, in the ZHZ stack, the bottom ZrO$_2$ layer can readily convert to ZrSiO$_4$ at 700 °C, significantly reducing the effective κ at the critical interface. In HZH, the bottom interface is HfO$_2$, which is less prone to silicate formation at that temperature, so the EOT penalty is less severe. Consequently, the detrimental impact of silicate formation is more pronounced in the ZHZ structure compared to HZH. Another factor contributing to increased leakage after annealing is grain growth in the polycrystalline HfO$_2$ and ZrO$_2$ layers. Prolonged or high temperature anneals can enlarge crystal grains and possibly form grain boundaries that serve as leakage pathways, slightly raising J$_g$.[28, 29]

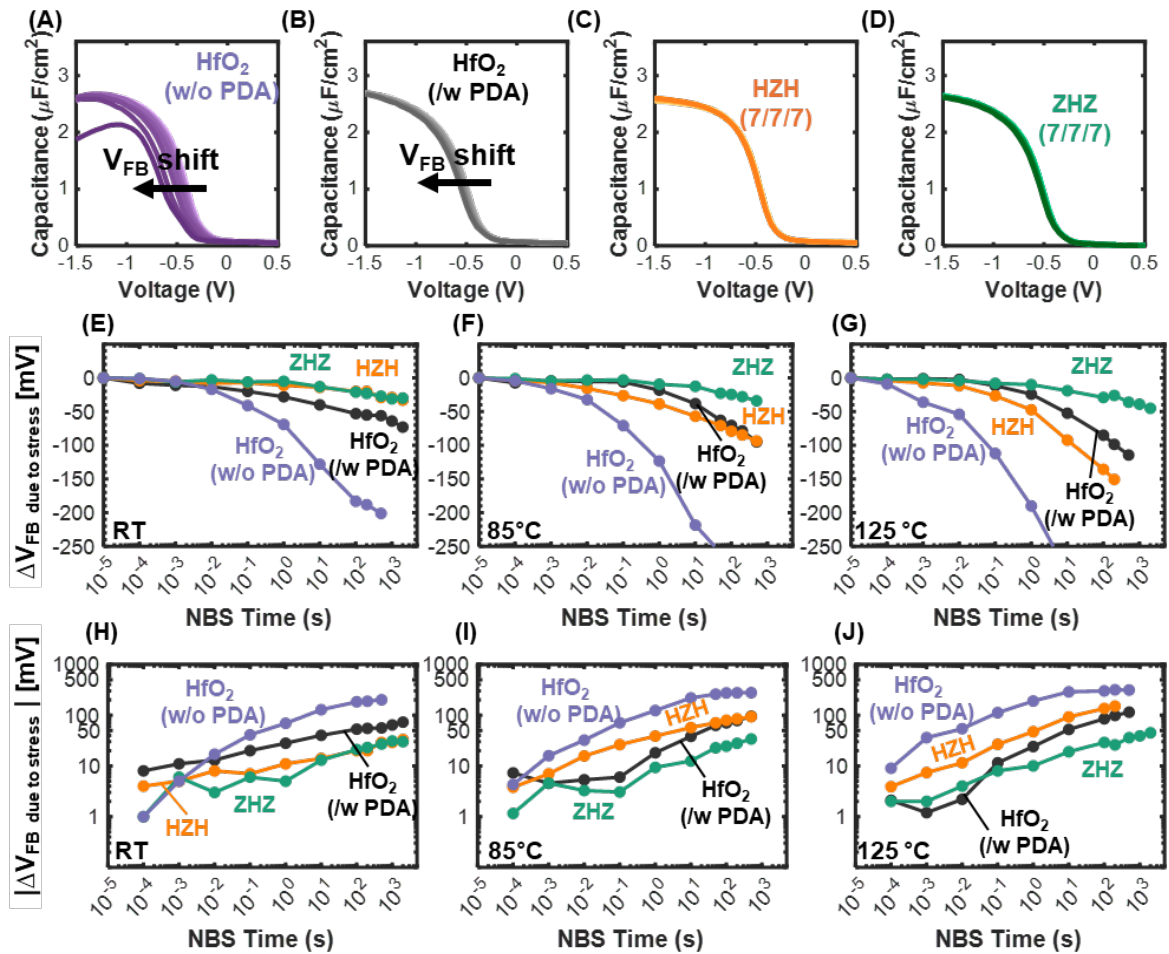

**Figure 3. NBS Measurement for Evaluating the Reliability of Various Gate Stacks**

(A)–(D) Capacitance-voltage (C–V) measurements under a negative bias (-2V) over a duration of 0 to 2000s. (E)–(G) Flatband voltage shift due to negative bias stress on a linear scale measured at (E) room temperature (RT), (F) 85 °C, and (G) 125 °C. (H)–(J) present the same data on a logarithmic scale at (H) RT, (I) 85 °C, and (J) 125 °C.

We next evaluated the NBS reliability of the gate stacks by measuring the flatband voltage shift under prolonged negative gate bias. MOS capacitors were stressed at –2 V for durations from 1 to 2000 s, and C–V curves were periodically measured to extract $V_{FB}$ (Figure 3A–D). The detailed NBS measurement pulse is illustrated in Figure S6. Tests were done at both room temperature (RT) and an elevated temperature of 85°C and 125 °C to assess thermal acceleration of instability. The unannealed $HfO_2$ sample showed a pronounced negative $V_{FB}$ shift of –182 mV after just 100 s of stress at RT, indicating substantial charge trapping in the dielectric (primarily hole trapping and interface trap generation).[30, 31] In contrast, the $HfO_2$ sample that received the 700 °C PDA exhibited a much smaller shift (–53 mV at 100 s, and –73 mV at 2000 s), highlighting that high-temperature annealing dramatically improves bias stability by reducing pre-existing defects.[32, 33] This underscores the necessity of PDA in mitigating charge trapping and ensuring long-term reliability of high-κ gate stacks.[32, 33]

The NBS characteristics further highlight the temperature-dependent reliability differences among the gate stacks. At room temperature, the HZH stack exhibits better NBS stability than the annealed $HfO_2$ control, while its performance becomes comparable at 85 °C and slightly worse at 125 °C. In contrast, the ZHZ stack maintains consistently superior reliability across all measured temperatures. At RT, the HZH and ZHZ samples exhibit only –33 mV and –30 mV shifts, respectively, after 2000 s of stress. At 125 °C, after 100 s of –2 V stress, annealed $HfO_2$ shows a –85 mV shift; on the other hand, the HZH and ZHZ stacks exhibit –135 mV and –29 mV shifts, respectively. Even after 2000 s at 125 °C, the ZHZ stack maintains the smallest degradation (–45 mV). These trends, summarized in Figure 3E–G, indicate that although PDA-treated samples exhibit broadly comparable stability at RT, the ZHZ stack clearly outperforms the others under high-temperature stress. Figures 3H–J present the same data on a logarithmic y-scale. The approximately linear behavior in the log–log plots confirms a power-law dependence of $\Delta V_{FB}$ on stress time, consistent with charge-trapping and interface-state generation mechanisms. The activation energies extracted from the Arrhenius analysis of $\Delta V_{FB}$ at 100 s (Figure S8) reveal a distinct asymmetry: HZH exhibits a substantially higher Ea whereas both annealed $HfO_2$ and the ZHZ superlattice show shallow activation energies. This divergence indicates asymmetric Hf↔Zr intermixing during high-temperature annealing—pronounced in HZH but strongly suppressed in ZHZ, where the $ZrO_2$ outer layers act as diffusion barriers. The deeper trap states formed through intermixing in HZH account for its stronger temperature acceleration, whereas the limited intermixing in ZHZ yields shallow trap kinetics and superior high-temperature reliability.

The superior robustness of ZHZ can be understood by examining its initial defect landscape and interfacial chemistry. Silicates such as $HfSiO_4$ and $ZrSiO_4$ are known to exhibit lower defect densities than their respective binary oxides.[34, 35] Because $ZrSiO_4$ forms at a lower temperature than $HfSiO_4$,[35-37] the ZHZ structure is more likely to develop a $ZrSiO_4$-rich transition layer during PDA. This interpretation is strongly supported by the Dit analysis in Figure S7: $ZrO_2$-containing stacks exhibit significantly lower Dit after PDA than their $HfO_2$ counterparts, particularly in the high-$\Delta E$ region. Thus, the ZHZ stack begins with a lower density of pre-existing interface traps, reducing the number of sites available for immediate occupation under negative bias. This lower initial trap density plays a decisive role in the early-time NBS response. With fewer traps available for rapid filling, the ZHZ stack shows much smaller $\Delta V_{FB}$ shifts during both short and long stress durations. The stress-bias dependence in Figure S9 further supports this conclusion: although $\Delta V_{FB}$ increases with both stress voltage and time, as expected for trap generation under higher fields, the ZHZ stack consistently exhibits the smallest $\Delta V_{FB}$ across all bias conditions. Overall, the combination of a lower initial trap density, favorable $ZrSiO_4$ interfacial formation, and a balanced superlattice arrangement results in the ZHZ gate stack demonstrating the strongest NBS reliability across all temperatures and stress conditions.

In summary, each gate stack configuration offers distinct advantages. The HZH structure provides enhanced capacitance compared to $HfO_2$ alone, making it appealing for applications that demand high drive current and performance. The ZHZ structure, while not improving EOT beyond the $HfO_2$ baseline, delivers superior reliability under stress, which is critical for applications requiring long-term stability. These findings underscore the importance of structural engineering in high-κ gate stacks to balance capacitance and leakage performance against reliability.

**Impact of Drive-in Temperature on Different Dipole Layer Locations**

To systematically study dipole layer placement, we fabricated devices with a 3 Å $Al_2O_3$ interfacial layer inserted at three different locations in an $HfO_2$-based gate stack: at the bottom interface (adjacent to the $SiO_2$ interfacial layer), in the middle of the $HfO_2$, or at the top of the $HfO_2$ (just below the metal gate). Figure 4A illustrates these configurations. Each of these samples was subjected to anneals at 700 °C, 800 °C, and 900 °C to examine how thermal budget influences dipole formation and activation. The flatband voltage shift ($\Delta V_{FB}$) due to the dipole was extracted from C–V curves for each case (Figure 4B–D).

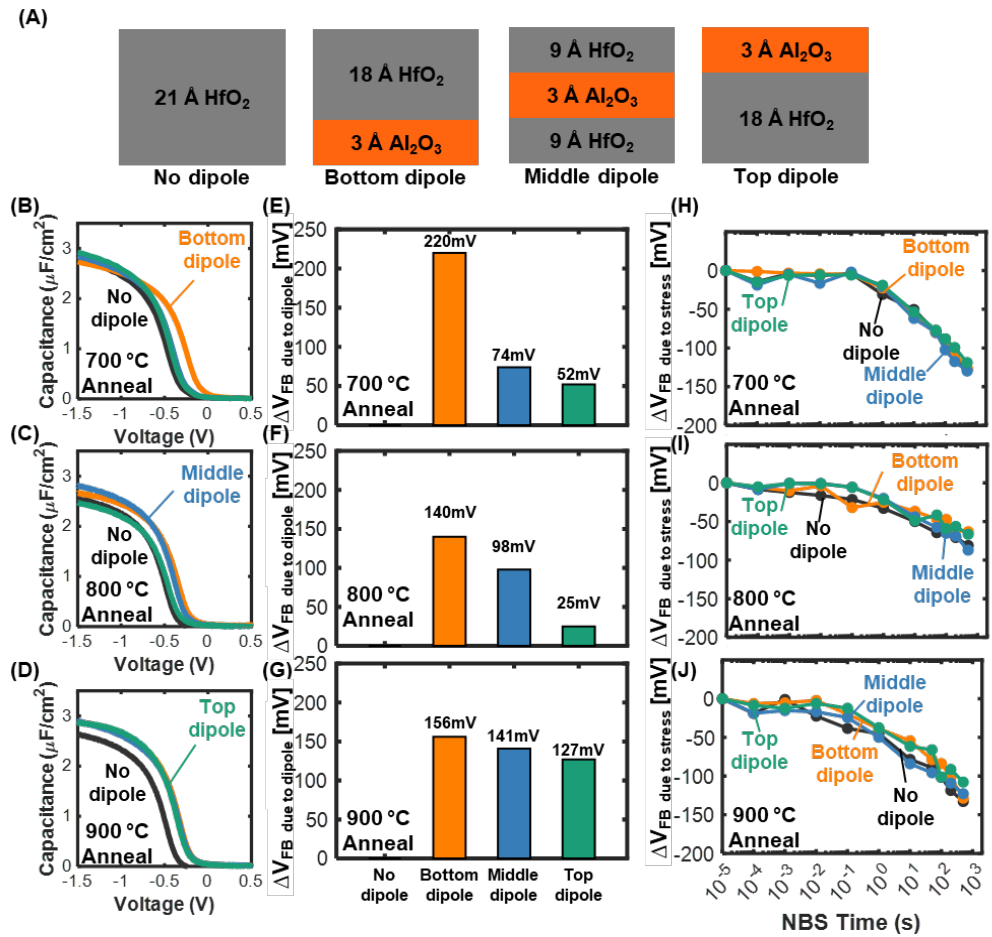

**Figure 4. Dipole engineering study with different Al$_2$O$_3$ locations and annealing temperatures**

(A) Schematic of gate stack configurations with different dipole locations (B)–(D) Capacitance–voltage (C–V) measurements of samples annealed at 700 °C, 800 °C, and 900 °C, respectively (E)–(G) Flatband voltage shift of samples annealed at 700 °C, 800 °C, and 900 °C, respectively (H)–(J) NBS measurement results at elevated temperature(125 °C) for samples annealed at 700 °C, 800 °C, and 900 °C, respectively

At 700 °C, a substantial V$_{FB}$ shift of +220 mV was observed only for the sample with the dipole at the bottom interface, whereas the mid-layer and top-layer dipole samples showed much smaller shifts of +74 mV and +52 mV, respectively (Figure 4E). This behavior indicates that at this relatively lower annealing temperature the dipole is fully activated only when positioned at the bottom interface, likely because it is directly adjacent to the SiO$_2$ interface where interfacial dipole formation is most favorable. The corresponding C–V curves (Figure 4B) confirm the large V$_{FB}$ shift for the bottom-dipole device and much less shifts for the other two.

The mechanism of dipole formation at oxide interfaces is rooted in differences in oxygen areal density between two adjoining layers. Generally, when a layer with a higher oxygen concentration is adjacent to one with a lower concentration, oxygen ions tend to diffuse into the lower concentration region upon annealing. This leaves behind oxygen vacancies in the higher concentration material. These oxygen vacancies (positively charged) at the interface align with oxygen anions (negatively charged) in the neighboring layer, thereby forming an electric dipole at the interface. The direction of the dipole depends on the direction of oxygen diffusion.[38]

In our experiments, we employed Al$_2$O$_3$ as the dipole-forming layer. When Al$_2$O$_3$ is placed at the bottom of the gate stack (directly on the SiO$_2$ interfacial layer), oxygen from Al$_2$O$_3$ diffuses into SiO$_2$ during annealing, creating vacancies in Al$_2$O$_3$. This produces dipoles oriented from the SiO$_2$ side toward the Al$_2$O$_3$, which induces a positive V$_{FB}$ shift.[39] By contrast, if the Al$_2$O$_3$ layer is located in the middle or top of the HfO$_2$ stack, it is initially separated from SiO$_2$ by HfO$_2$. Upon annealing, some Al and O interdiffusion can occur through the HfO$_2$, and dipoles can still form at the nearest SiO$_2$ interface. However, the farther the Al$_2$O$_3$ is from the SiO$_2$, the less oxygen exchange occurs, resulting in a much smaller dipole and hence a smaller V$_{FB}$ shift. This explains why at 700 °C only the bottom dipole showed a significant effect, whereas the middle and top dipoles did not.

When the annealing temperature increased to 800 °C, the dipole in the middle of the stack became more activated, showing a larger Δ V$_{FB}$ of 98 mV, while the bottom dipole's effect slightly decreased to 140 mV (Figure 4F). At 900 °C, the top dipole configuration exhibited a pronounced $V_{FB}$ shift (127 mV), nearly catching up to Δ $V_{FB}$ of 156 mV with the bottom dipole structure which remained roughly constant between 800 °C and 900 °C as shown in Figure 4G. These trends suggest that higher annealing temperatures enhance the diffusion of oxygen and subsequent dipole formation even for dipoles that are initially farther from the $SiO_2$ interface. In other words, a sufficiently high-temperature "drive-in" can activate dipole formation throughout the high-κ stack, not just at the bottom interface. To activate dipoles located away from the immediate interface, we used a higher-temperature drive-in anneal (up to 900 °C). At such high temperatures, the diffusion lengths of oxygen in $HfO_2$ increase substantially, enabling oxygen from $Al_2O_3$ even in the middle or upper part of the stack to reach the $SiO_2$ eventually, thereby forming measurable dipoles (Figure 4G).[24, 40]

Conversely, for the bottom dipole, as the net dipole effect at the bottom interface slightly diminishes beyond 700 °C, there appears to be an optimal annealing window. An intriguing behavior was observed for the bottom dipole case: The $V_{FB}$ shift caused by the bottom $Al_2O_3$ layer was largest at 700 °C and then reduced at 800 °C and 900 °C. This non-monotonic trend can be understood by considering the so-called mirror-plane effect in high-κ dielectrics. In a $HfO_2$-based dielectric stack, there can be a self-regulating interdiffusion of cations and anions at interfaces that tends to neutralize dipoles.[25] At moderate temperatures, enough oxygen diffuses to form a strong dipole at the $SiO_2/Al_2O_3$ interface, while Hf and other cation diffusion is limited. At much higher temperatures, however, oxygen diffusion becomes extremely pronounced and can extend beyond the first interface: in our case, oxygen from $Al_2O_3$ can diffuse not only into $SiO_2$ but also upward into the adjacent $HfO_2$ layer. Concurrently, Hf cations can diffuse into the $Al_2O_3$ region to some extent. This upward anion/cation intermixing generates an opposite interfacial dipole component—i.e., a counter-dipole—that partially cancels the intended dipole at the $SiO_2$ interface. Specifically, since $HfO_2$ has a lower oxygen areal density than $Al_2O_3$, oxygen will also diffuse from $Al_2O_3$ into $HfO_2$ at high temperature, creating vacancies in $Al_2O_3$ near the $HfO_2$ side. This induces an opposing dipole oriented from $Al_2O_3$ toward $HfO_2$.[38, 41] At 900 °C, this counter-dipole effect remains present but is not strictly monotonic: the net $V_{FB}$ shift is still lower than at 700 °C but slightly higher than at 800 °C, consistent with temperature-dependent competition between dipole formation and counter-dipole cancellation. In summary, these results demonstrate that excessive high-temperature diffusion introduces a counter-dipole that reduces the net dipole strength, and a careful control of the annealing temperature is required to maximize the beneficial dipole at the $SiO_2$ interface while avoiding excessive counter-dipole formation due to over-diffusion.

We further evaluated the dipole-engineered stacks under negative bias stress (NBS) across multiple temperatures, including room temperature (RT), 85 °C, and 125 °C. The elevated-temperature results at 125 °C are presented in Figure 4H–J, while the corresponding RT and 85 °C data are provided in Figure S10. The NBS pulse conditions were consistent with those described in Figure S6. As expected, applying –2 V stress induced a negative shift in $V_{FB}$ in all cases due to charge trapping. Importantly, however, the magnitude of the $V_{FB}$ shift after stress was similar for devices with bottom, middle, or top dipoles and largely independent of whether they were annealed at 700, 800, or 900 °C. In other words, within the measurement resolution, dipole layer placement and the drive-in anneal temperature did not significantly affect the NBS-induced degradation. All devices showed only modest $V_{FB}$ shifts under prolonged stress, indicating that the introduction of the $Al_2O_3$ dipole layer does not compromise the reliability of the gate stack.

The impact of dipole insertion on channel mobility and reliability is an important consideration for practical devices. Some studies have reported that if a dipole-forming layer such as $Al_2O_3$ is directly interfaced with $SiO_2$, it can lead to mobility degradation in MOSFETs, possibly due to scattering from the dipole interface.[24] Conversely, other reports suggest that dipoles can improve reliability by providing a source of excess oxygen to passivate interface traps, thereby reducing interface state density ($D_{it}$).[42] Our reliability tests on capacitors with various dipole placements (Figure 4H–J) showed no adverse effect of the dipole on NBS-induced $V_{FB}$ shifts even at elevated temperature(125°C) – all configurations were essentially equally stable, and in fact the absolute shifts were very small. This suggests that, at least in terms of bias stress, adding an $Al_2O_3$ dipole layer does not degrade the dielectric reliability. It will be necessary, however, to conduct bias-temperature instability (BTI) tests on actual transistors to fully evaluate any subtle effects dipoles might have on channel mobility or interface trap generation over operating lifetimes.

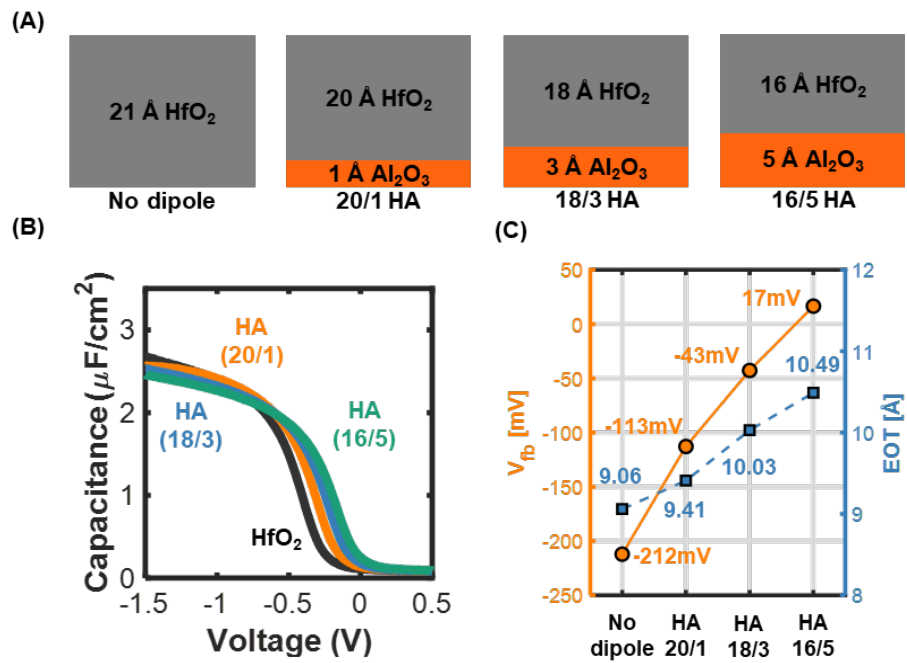

**Figure 5. Demonstration of the EOT–$V_{FB}$ shift trade-off by modulating $Al_2O_3$ dipole thickness in $HfO_2/Al_2O_3$ gate stacks.** (A) Schematic of gate stack configurations with different dipole thicknesses (B) Capacitance–voltage (C–V) measurements of HA stacks with varying $Al_2O_3$ thicknesses (C) Flatband voltage ($V_{FB}$) and equivalent oxide thickness (EOT) values as a function of $Al_2O_3$ dipole thickness

To evaluate threshold voltage tunability, we investigated the dependence of electrical characteristics on $Al_2O_3$ thickness in the gate stack structures, as illustrated in Figure 5A. Three different $Al_2O_3$ thicknesses, 1 Å, 3 Å, and 5 Å were incorporated by partially replacing the 21 Å $HfO_2$ layer, ensuring that the total dielectric thickness remained constant across all samples. This approach allowed for a controlled assessment of dipole strength while minimizing capacitance variations caused by total thickness changes. The capacitance–voltage (C–V) measurements (Figure 5B) show a clear positive shift of the C–V curve as the $Al_2O_3$ thickness increases, accompanied by a decrease in maximum capacitance. The extracted $V_{FB}$ values exhibit a strong correlation with $Al_2O_3$ thickness (Figure 5C): +99 mV for 1 Å, +169 mV for 3 Å, and +229 mV for 5 Å, confirming that the dipole effect intensifies with increasing $Al_2O_3$ thickness. This enhancement is attributed to interfacial dipole formation, driven by differences in oxygen areal density between adjacent dielectric layers. Thicker $Al_2O_3$ layers promote greater oxygen diffusion, resulting in a stronger dipole moment and a larger positive shift in $V_{FB}$. The EOT values, extracted from the C–V curves (Figure 5C), show a clear degradation trend with increasing $Al_2O_3$ thickness due to its intrinsically lower dielectric constant compared to $HfO_2$. Specifically, EOT increased by 0.35 Å, 0.97 Å, and 1.43 Å for 1 Å, 3 Å, and 5 Å of $Al_2O_3$, respectively. These results clearly demonstrate the intrinsic EOT–$V_{FB}$ trade-off in conventional $HfO_2$/$Al_2O_3$ dipole stacks, motivating the use of intrinsically low-EOT platforms—such as the Hf/Zr superlattice—to enable meaningful $V_{th}$ modulation without compromising scaling. In this context, it modulation integrates with low-EOT superlattice structures such as HZH and to determine the is important to evaluate how dipole-based $V_{th}$ implications for advanced CMOS design.

**$V_{FB}$ Modulation in Superlattice High-κ Gate Stacks through Dipole Engineering**

To investigate the effect of incorporating an Al$_2$O$_3$ dipole into previously examined gate stack structures, we embedded the dipole layer into HfO$_2$/ZrO$_2$ superlattice configurations. By doing so, we leverage the enhanced permittivity of the laminate while gaining the V$_{FB}$ modulation of the dipole, thus balancing electrostatic control with capacitance retention. We extended our investigation beyond the HfO$_2$-only dielectric to three laminated high-κ stacks (HZH, ZHZ, and HZZ), each incorporating a bottom-interface Al$_2$O$_3$ dipole (Figure 6A).

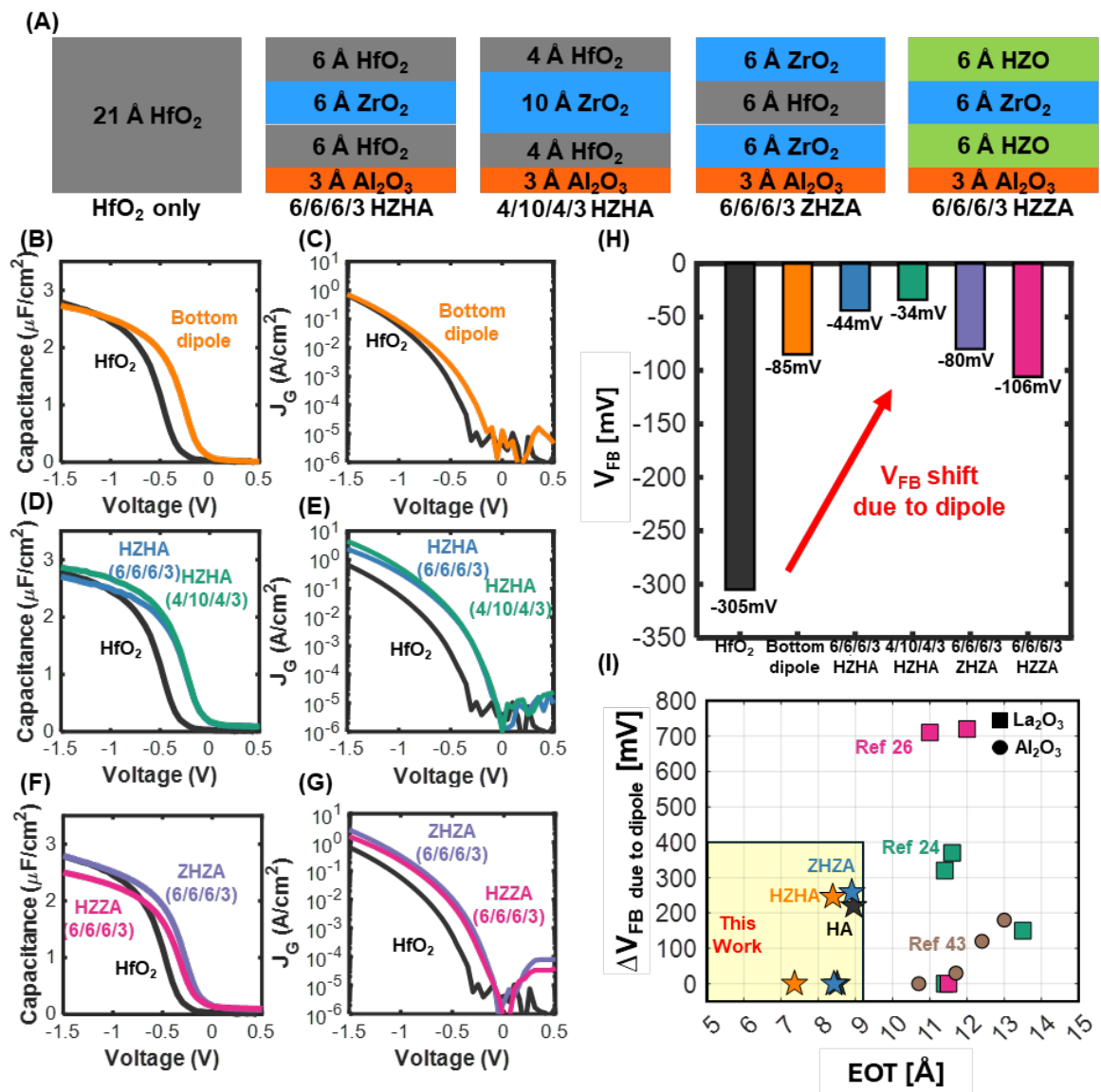

**Figure 6. Incorporation of a dipole into the Hf/Zr Superlattice structure for Vth tunability and enhanced high-k performance** (A) Schematic of the $HfO_2$-only, HZH, ZHZ, and HZZ gate stack configurations with the dipole layer positioned at the bottom. (B) and (C) Capacitance–voltage (C–V) and gate leakage current–voltage ($J_g$–V) characteristics for $HfO_2$ and $HfO_2$ with a bottom dipole. (D) and (E) Capacitance–voltage (C–V) and gate leakage current–voltage ($J_g$–V) characteristics for $HfO_2$ and HZH gate stacks with a bottom dipole. (F) and (G) Capacitance–voltage (C–V) and gate leakage current–voltage ($J_g$–V) characteristics for $HfO_2$ and ZHZ/HZZ gate stacks with a bottom dipole. (H) Flatband voltage of different gate stacks. Regardless of the gate stack configuration, those incorporating a bottom dipole exhibit a flatband voltage shift. (I) Benchmark comparison of the studied gate stacks with results from other studies.

Figure 6B–G presents C–V and $J_g$–V data for several representative cases with a bottom dipole. The introduction of the 3 Å $Al_2O_3$ layer at the $HfO_2/SiO_2$ interface consistently induces a positive flatband voltage shift, confirming the formation of the intended dipole. Quantitatively, adding the dipole to a pure $HfO_2$ gate stack (sample "HA," referring to $HfO_2$ + bottom dipole $Al_2O_3$) shifted $V_{FB}$ by about +220 mV relative to the undoped $HfO_2$, but also increased the EOT from 8.5 Å to 9.0 Å. Importantly, the gate leakage current density remained essentially unchanged by the dipole (Figure 6B, C). This demonstrates that the $Al_2O_3$ interlayer can modulate $V_{th}$ without incurring a severe leakage penalty, although a slight capacitance reduction is observed.

The stack composition with the dipole significantly influences the overall device behavior. Among the high-κ laminates tested, the HZH stack showed the lowest EOT even with the dipole in place—even slightly lower than the conventional $HfO_2$ without a dipole. In contrast, the ZHZ stack with a dipole exhibited a lower capacitance and higher leakage compared to its no-dipole counterpart, underlining that the base $HfO_2/ZrO_2$ configuration can matter as much as the dipole itself. We further examined two HZH stack variants with different $HfO_2/ZrO_2$ thickness ratios to see how they interact with the dipole. Without a dipole, the symmetric 7/7/7 HZH had a lower EOT than the asymmetric 4/13/4 HZH, as discussed earlier. When the dipole was added at the bottom interface, an interesting result emerged: the 4/10/4/3 HZHA stack achieved a lower EOT (8.4 Å) than the 6/6/6/3 HZHA stack (9.2 Å). In other words, the laminate with a higher $ZrO_2$ proportion benefitted more from the dipole integration in terms of maintaining a low EOT. By contrast, other dipole-integrated configurations like ZHZA ($ZrO_2/HfO_2/ZrO_2$ + $Al_2O_3$) and HZZA (HZO/$ZrO_2$/HZO + $Al_2O_3$) exhibited significantly higher EOTs (8.9 Å and 10.0 Å, respectively) despite the dipole, due to their less favorable base structures. These findings emphasize that to fully exploit dipole engineering, one must jointly optimize the high-κ stack geometry. A well-tuned $HfO_2$:$ZrO_2$ ratio can offset the intrinsic EOT cost of the $Al_2O_3$ layer, thereby maximizing capacitance while still enabling $V_{FB}$ shifts.

We calculated the flatband voltages for all the different gate stacks to compare the effect of the dipole (Figure 6H). In each case, incorporating the $Al_2O_3$ dipole resulted in a discernible $V_{FB}$ shift relative to the corresponding structure without a dipole. The magnitude of the shift varied depending on the stack composition (with HZH showing the largest shift and HZZ the smallest), but the qualitative presence of a shift in HZH, ZHZ, and HZZ confirms successful dipole formation in all these high-κ matrices. This demonstrates the versatility of dipole engineering across various dielectric heterostructures; even when the baseline $V_{FB}$ of HZH, ZHZ, and pure $HfO_2$ devices are nearly identical, adding the dipole reliably moves $V_{FB}$ in the positive direction.

Figure 6I provides a performance benchmarking of our dipole-integrated stacks in terms of flatband voltage shift($\triangle V_{FB}$) versus EOT.[43] As noted, the $HfO_2$-only + dipole stack

(HA) incurs a slight EOT increase compared to pure $HfO_2$. In contrast, the HZH + dipole stack manages to achieve a lower EOT than pure $HfO_2$ due to the $ZrO_2$ layers while still obtaining a sizable $V_{FB}$ shift from the dipole. This underlines the promise of combining laminate high-κ dielectrics with dipole layers: the laminate addresses the capacitance requirement, and the dipole provides $V_{th}$ tuning, with manageable trade-offs.

Figure 6I benchmarks the flat-band voltage tunability ($\Delta V_{FB}$) against equivalent oxide thickness for various dipole-engineered gate stacks reported in the literature. Prior studies employing $La_2O_3$ or $Al_2O_3$ interlayers have demonstrated substantial $\Delta V_{FB}$ (>300–700 mV) but at the cost of large EOTs (20–50 Å). In contrast, our $HfO_2$–$ZrO_2$ superlattice platform enables the integration of a 3 Å $Al_2O_3$ dipole within sub-10 Å EOT stacks: the HZHA and ZHZA structures both exhibit $\Delta V_{FB}$ shifts in excess of 200 mV while maintaining EOTs of 8.4 Å and 9.0 Å, respectively. This performance region (highlighted in yellow) lies well below the EOT regime of previous dipole-only approaches, illustrating that coupling dipole engineering with high-κ superlattice dielectric stacks is a highly effective strategy to achieve aggressive scaling and threshold-voltage tunability simultaneously.

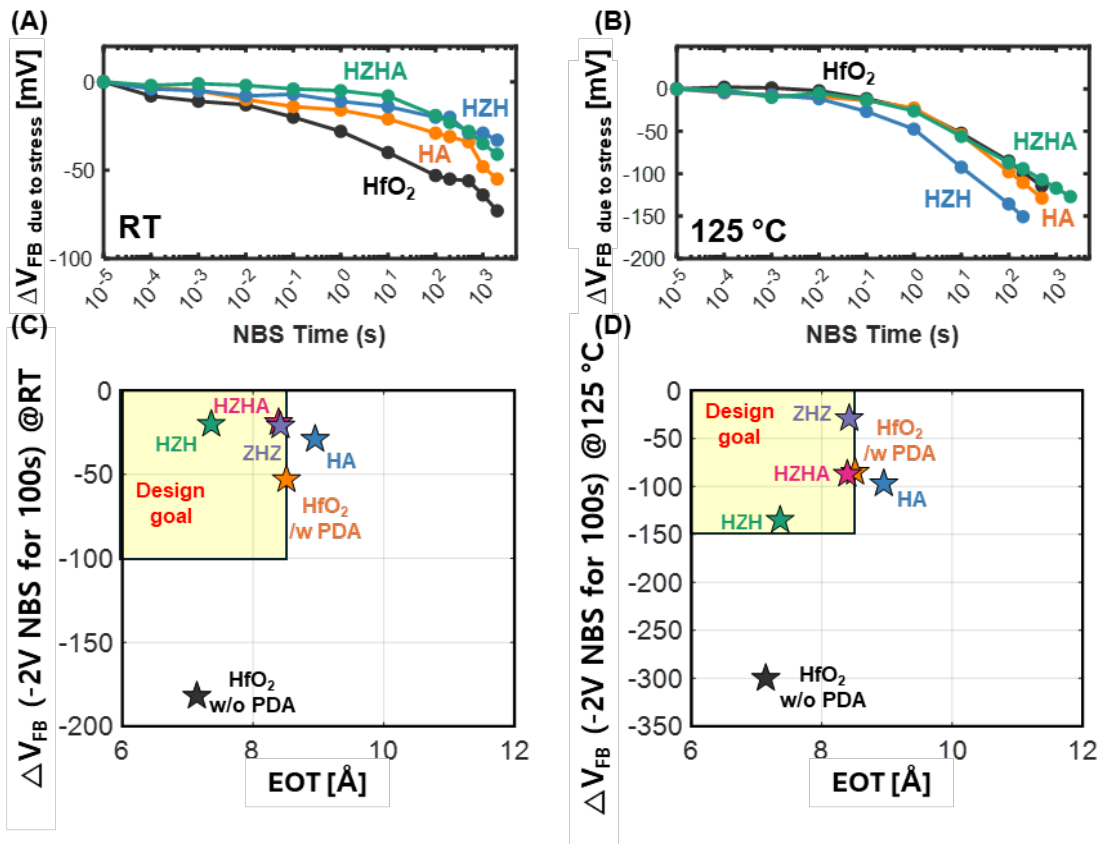

**Figure 7. NBS measurements of gate stacks with incorporated dipole layer**

(A) and (B) Flatband voltage shift ($\triangle V_{FB}$) as a function of negative bias stress (NBS) time measured at room temperature and 125 °C for various gate stack configurations. (C) and (D) Flatband voltage shift ($V_{FB}$) versus EOT plot for the structures investigated in this work.

We further evaluated the NBS reliability of these dipole-integrated superlattice stacks. Figure 7A–B compares the NBS response of four stacks—$HfO_2$, HA, HZH, and HZHA—at room temperature and 125 °C. At both temperatures, introducing an ultrathin $Al_2O_3$ dipole directly on the $SiO_2$ interlayer (HA) does not degrade reliability; the $\Delta V_{FB}$ versus time trace remains similar to, or modestly better than, that of the conventional $HfO_2$. This behavior suggests that the dipole interface supplies excess oxygen during stress, promoting partial conversion of $HfO_2$ to Hf–Al silicate and thereby passivating vacancy sites that would otherwise trap holes. When the dipole is embedded within the $HfO_2/ZrO_2/HfO_2$ superlattice (HZHA), the stack retains the superlattice's intrinsic NBS robustness at room temperature and, importantly, sustains comparable or slightly smaller $\Delta V_{FB}$ than conventional $HfO_2$ at 125 °C. The combination of Zr-mediated trap suppression and Al-assisted interfacial silicate formation therefore enables the HZHA architecture to deliver threshold voltage tunability without sacrificing high temperature reliability, affirming its suitability for advanced CMOS logic applications. Figure 7C–D plots the 100 s negative-bias stress $\Delta V_{FB}$ against EOT for all investigated stacks at room temperature and 125 °C respectively. The shaded "design goal" region denotes the quadrant combining EOT < 8.5 Å with $|V_{FB}|$ < 100 mV. At both temperatures, the HZHA stack lies within this target region, exhibiting an EOT of 8.4 Å alongside minimal $\Delta V_{FB}$ (< 20 mV at RT and < 100 mV at 125 °C). For comparison, the HA stack provides tunability but falls outside the design window because of its larger EOT, whereas the conventional $HfO_2$ stack without PDA exhibits a substantial flatband voltage shift. Notably, HZHA maintains low $\Delta V_{FB}$ under high-temperature stress—demonstrating that embedding the dipole within the $HfO_2$–$ZrO_2$ superlattice uniquely satisfies the competing requirements of sub-nanometer scaling and robust negative-bias reliability. In parallel, our NBS data highlight the strong potential of the ZHZ architecture for applications requiring enhanced reliability under both thermal and electrical stress. By naturally forming a more inert interface and maintaining a high overall $\kappa$, the ZHZ stack achieves an excellent balance of performance and stability. It points to a broader strategy where interface chemistry is engineered for reliability in tandem with bulk high-$\kappa$ improvements.

Finally, by incorporating an $Al_2O_3$ dipole into our best-performing high-$\kappa$ laminates (HZH, ZHZ, HZZ), we demonstrated clear flatband voltage modulation in each case (Figure 6H). The fact that the baseline $V_{FB}$ of HZH, ZHZ, and HZZ without dipole could be made nearly identical through appropriate metal gate work function, and then each could be shifted positively by adding the dipole, confirms the viability of dipole engineering across different high-κ systems. The dipole did incur a minor EOT due to a lower dielectric constant of $Al_2O_3$ than that of $HfO_2$ and $ZrO_2$,[22, 23] but the leakage current remained largely unchanged. Notably, the HZHA (4/10/4/3) stack achieved a slightly lower EOT than even some no-dipole configurations, illustrating that a well-optimized $HfO_2/ZrO_2$ composition can compensate for the $Al_2O_3$'s lower κ. This implies that by tuning the Hf:Zr ratio in concert with dipole integration, one can simultaneously maximize the dipole effect and minimize the EOT penalty, ultimately preserving or even enhancing the overall capacitance.

Table 1 synthesizes the key performance metrics of the investigated gate stacks—EOT, gate leakage density, flatband voltage shifts after negative bias stress at both room temperature and 125 °C, $V_{th}$ tunability, and compatibility with replacement metal gate (RMG) processing. Relative to the $HfO_2$ control, HZH superlattices deliver thinner EOTs without sacrificing significant leakage integrity, and the 700 °C post-deposition anneal substantially suppresses NBS-induced $V_{FB}$ drift. Incorporating an ultrathin $Al_2O_3$ dipole layer increases EOT for any stack; however, the high-κ HZH framework offsets this penalty, such that the dipole-integrated HZHA stack retains competitive EOT while gaining intrinsic $V_{th}$ adjustability. Moreover, HZHA maintains NBS stability at elevated temperature and endures the high temperature anneal, confirming full compatibility with advanced RMG flows. Collectively, the data position HZHA as the most balanced candidate for future logic nodes, combining aggressive scaling, threshold voltage control, and robust reliability.

| Gate stack | HfO$_2$ only w/o PDA | HfO$_2$ only /w PDA | HZH w/o PDA | HZH /w PDA | ZHZ /w PDA | HA /w PDA | HZHA /w PDA |
|---|---|---|---|---|---|---|---|
| | HfO$_2$ | HfO$_2$ | HfO$_2$/ZrO$_2$/HfO$_2$ | HfO$_2$/ZrO$_2$/HfO$_2$ | ZrO$_2$/HfO$_2$/ZrO$_2$ | HfO$_2$/Al$_2$O$_3$ | HfO$_2$/ZrO$_2$/HfO$_2$/Al$_2$O$_3$ |
| EOT [Å] | 7.1 | 8.5 | 7.0 | 7.3 | 8.4 | 9.0 | 8.4 |
| J$_g$ @V$_{FB}$ -1V [A/cm$^2$] | 0.20 | 0.31 | 0.28 | 2.36 | 0.59 | 0.64 | 0.68 |
| V$_{FB}$ shift due to NBS @100s at RT [mV] | -182 | -53 | N/A | -20 | -21 | -29 | -19 |
| V$_{FB}$ shift due to NBS @100s at 125°C [mV] | -300 | -85 | N/A | -135 | -29 | -97 | -87 |
| V$_{th}$ tunability | X | X | X | X | X | ✓ | ✓ |
| RMG process compatibility | X | ✓ | X | ✓ | ✓ | ✓ | ✓ |

**Table 1. Summary table of characteristics for various types of gate stacks**

Table 2 benchmarks MFIS gate-stack performance reported in prior studies, comparing annealing conditions, EOT/CET, and V$_{th}$ tunability. The color coding indicates CMOS process compatibility, with green representing the most favorable conditions, yellow indicating intermediate performance, and red indicating the least favorable outcome. Notably, the gate dielectric developed in this work withstands high-temperature annealing up to 700 °C while still achieving a low EOT enabled by the HfO$_2$/ZrO$_2$ superlattice structure. Furthermore, the incorporation of an ultrathin Al$_2$O$_3$ dipole layer within the superlattice allows controllable

$V_{FB}$ tuning, demonstrating that both aggressive EOT scaling and threshold-voltage engineering can be simultaneously realized within a CMOS-compatible thermal budget.

| Ref. | Structure $T_{ox}$ | PDA (post-deposition annealing) | | | PMA (post-metallization annealing) | | | EOT [Å] | CET [Å] | Vth tunability |
|---|---|---|---|---|---|---|---|---|---|---|
| | | Temp | Time | Ambient | Temp | Time | Ambient | | | |
| **This Work** | HZH 21 Å | 700 °C | 30 s | N2 | - | | | /w PDA 7.3 w/o PDA 7.0 | - | ✓ |
| 13 | HZH 20 Å | 175 °C | 20 m | FGA | 175 °C | 20 m | FGA | 6.5 | - | X |
| 16 | HZO 25 Å | - | | | 600 °C | spike | - | - | 12 | X |
| 44 | HZO 15 Å | - | | | 450 °C | - | - | 12.5 | - | X |
| 45 | HZO 60 Å | - | | | 500 °C | 30 s | - | - | - | X |
| 46 | HZO 100 Å | - | | | 500 °C | 30 s | N2 | - | - | X |
| 47 | HZO 50 Å | - | | | 600 °C | 30 s | Ar | - | 9.8 | X |
| 48 | HZO 24 Å | 250 °C | - | - | 450 °C | 60 s | N2 | 7.1 | - | X |

**Table 2. Benchmark table of MFIS gate-stack performance against state-of-the-art works, comparing annealing conditions, EOT/CET, and $V_{th}$ tunability.** Color coding qualitatively benchmarks CMOS compatibility: green is most favorable, yellow is intermediate, and red is least favorable.

**CONCLUSION**

In this work, we demonstrate that HfO$_2$/ZrO$_2$ superlattice gate stacks incorporating ultrathin Al$_2$O$_3$ dipole layers simultaneously achieve aggressive EOT scaling, low leakage, tunable flat-band voltage, and high-temperature CMOS compatibility. The optimized HZH laminate delivers a 7.3 Å EOT—well below the 8.5 Å of conventional HfO$_2$—while dipole integration enables >200 mV $V_{FB}$ modulation with an 8.4 Å EOT in the HZHA stack. Although ZHZ does not further reduce EOT, it exhibits outstanding negative-bias stress resilience, reducing the 125 °C $\Delta V_{FB}$ from 85 mV (HfO$_2$) to 29 mV. In addition to these device-level advances, the study reveals two previously unreported physical insights—position-dependent dipole activation with a high-temperature counter-dipole response, and asymmetric Hf↔Zr intermixing that dictates trap energetics—offering new understanding of dipole formation, ionic diffusion, and defect behavior in ultrathin high-κ heterostructures. Collectively, these results position dipole-engineered $HfO_2$/$ZrO_2$ superlattices as a promising pathway toward sub-nanometer, high-performance gate dielectrics for future CMOS nodes.

## METHOD

The detailed device fabrication process is described in Note S1, while the device measurement methods for C-V, $J_g$-V, and NBS are explained in Note S2. Additionally, the parameter extraction method for $V_{FB}$, EOT, and $J_g$ at $V_{FB}$ - 1V is provided in Note S3.

## AUTHOR INFORMATION

**Corresponding Author**

*T.S. E-mail: tsong77@gatech.edu

* S.K. E-mail: skang415@gatech.edu

**Author Contributions**

A.K., and C.S. supervised the research. T.S., S.K., and Y.K. designed the research. T.S., S.K., and Y.K. conducted the experiments and analyzed the results. Y.K., J.C., L.F., N.A., and M.T. assisted with the data analysis. T.S., and S.K. wrote the manuscript with input from all authors.

## ACKNOWLEDGMENT

This work was supported by SRC GRC 2024 Logic and Memory Devices (#3235.001) and the National Research Foundation of Korea (NRF) grant funded by the Korea government (MSIT)

(No. RS-2023-00260527). We also acknowledge Seunghoon Sung, Ashish Penumatcha, and Uygar Avci of Intel Foundry Technology Research (Hillsboro, OR, USA, and Ireland) for their valuable contributions and insightful discussions.## REFERENCES

(1) Loubet, N.; Hook, T.; Montanini, P.; Yeung, C. W.; Kanakasabapathy, S.; Guillorn, M.; Yamashita, T.; Zhang, J.; Miao, X.; Wang, J.; et al. Stacked nanosheet gate-all-around transistor to enable scaling beyond FinFET. In 2017 Symposium on VLSI Technology, 2017.

(2) Ajayan, J.; Nirmal, D.; Tayal, S.; Bhattacharya, S.; Arivazhagan, L.; Fletcher, A. S. A.; Murugapandiyan, P.; Ajitha, D. Nanosheet field effect transistors-A next generation device to keep Moore's law alive: An intensive study. *Microelectronics Journal* **2021**, *114*. DOI: 10.1016/j.mejo.2021.105141.

(3) Zhang, Q.; Zhang, Y.; Luo, Y.; Yin, H. New structure transistors for advanced technology node CMOS ICs. *Natl Sci Rev* **2024**, *11* (3), nwae008. DOI: 10.1093/nsr/nwae008  From NLM PubMed-not-MEDLINE.

(4) Huang, J.; Heh, D.; Sivasubramani, P.; Kirsch, P.; Bersuker, G.; Gilmer, D.; Quevedo-Lopez, M.; Hussain, M.; Majhi, P.; Lysaght, P. Gate first high-k/metal gate stacks with zero SiO x interface achieving EOT= 0.59 nm for 16nm application. In *2009 Symposium on VLSI Technology*, 2009; IEEE: pp 34-35.

(5) Changhwan, C.; Chang Yong, K.; Se Jong, R.; Abkar, M. S.; Krishna, S. A.; Manhong, Z.; Hyungseob, K.; Tackhwi, L.; Feng, Z.; Injo, O.; et al. Fabrication of TaN-gated ultra-thin mosfets (eot >1.0nm) with HfO2 using a novel oxygen scavenging process for sub 65nm application. In Digest of Technical Papers. 2005 Symposium on VLSI Technology, 2005., 2005.

(6) Ando, T.; Frank, M. M.; Choi, K.; Choi, C.; Bruley, J.; Hopstaken, M.; Copel, M.; Cartier, E.; Kerber, A.; Callegari, A.; et al. Understanding mobility mechanisms in extremely scaled HfO2 (EOT 0.42 nm) using remote interfacial layer scavenging technique and Vt-tuning dipoles with gate-first process. In 2009 IEEE International Electron Devices Meeting (IEDM), 2009.

(7) Bao, R.; Durfee, C.; Zhang, J.; Qin, L.; Rozen, J.; Zhou, H.; Li, J.; Mukesh, S.; Pancharatnam, S.; Zhao, K.; et al. Critical Elements for Next Generation High Performance Computing Nanosheet Technology. In 2021 IEEE International Electron Devices Meeting (IEDM), 2021.

(8) Bao, R.; Watanabe, K.; Zhang, J.; Guo, J.; Zhou, H.; Gaul, A.; Sankarapandian, M.; Li, J.; Hubbard, A. R.; Vega, R.; et al. Multiple-Vt Solutions in Nanosheet Technology for High Performance and Low Power Applications. In 2019 IEEE International Electron Devices Meeting (IEDM), 2019.

(9) Zhang, J.; Ando, T.; Yeung, C. W.; Wang, M.; Kwon, O.; Galatage, R.; Chao, R.; Loubet, N.; Moon, B. K.; Bao, R.; et al. High-k metal gate fundamental learning and multi-Vt options for stacked nanosheet gate-all-around transistor. In 2017 IEEE International Electron Devices Meeting (IEDM), 2017.

(10) Bao, R.; Zhou, H.; Wang, M.; Guo, D.; Haran, B. S.; Narayanan, V.; Divakaruni, R. Extendable and Manufacturable Volume-less Multi-Vt Solution for 7nm Technology Node and Beyond. In 2018 IEEE International Electron Devices Meeting (IEDM), 2018.

(11) Yao, J.; Wei, Y.; Yang, S.; Yang, H.; Xu, G.; Zhang, Y.; Cao, L.; Zhang, X.; Liu, Q.; Wu, Z.; et al. Record 7(N)+7(P) Multiple VTs Demonstration on GAA Si Nanosheet n/pFETs using